# Improving thermoelectric performance of TiNiSn by mixing MnNiSb in the half-Heusler structure

T. Berry[a], S. Ouardi[b], G. H. Fecher[a], B. Balke[b], G. Kreiner[a], G. Auffermann[a], W. Schnelle[a], and C. Felser[a]

The thermoelectric properties of n-type semiconductor, TiNiSn is optimized by partial substitution with metallic, MnNiSb in the half Heusler structure. Herein, we study the transport properties and intrinsic phase separation in the $Ti_{1-x}Mn_xNiSn_{1-x}Sb_x$ system. The alloys were prepared by arc-melting and were annealed at temperatures obtained from differential thermal analysis and differential scanning calorimetry results. The phases were characterized using powder X-ray diffraction patterns, energy-dispersive X-ray spectroscopy, and differential scanning calorimetry. After annealing the majority phase was TiNiSn with some Ni-rich sites and the minority phases was majorly $Ti_6Sn_5$, Sn, and $MnSn_2$. Ni-rich sites were caused by Frenkel defects, this led to a metal-like behavior of the semiconducting specimens at low temperature. For x≤0.05 the samples showed an activated conduction, whereas for x>0.05 they showed metallic character. The figure of merit for x=0.05 was increased by 61% (ZT=0.45) in comparison to the pure TiNiSn.

## Introduction

Energy usage optimization and sustainability have been a research goal for decades. Amongst other avenues to contribute to this effort of sustainable energy, thermoelectric (TE) materials are of particular interest in solid state chemistry. The waste heat energy given out by power generating devices can be used to internalize as electrical energy. The efficiency of non-magnetic TE material can be determined by the figure of merit, *zT*,

$$zT = \frac{\sigma \alpha^2}{\kappa_e + \kappa_l} T$$

where $\alpha$ is the Seebeck coefficient, σ is the electrical conductivity and $\kappa_e$ and $\kappa_l$ is the electronic and lattice thermal conductivity, respectively.

TiNiSn-type TE materials are widely studied because of their relatively high Seebeck coefficient, electrical conductivity, and mechanical stability.[1] Various attempts to improve the performance of TiNiSn have been made, such as isoelectronic substitution of Ti by Hf and/or Zr, p-type doping by Sc, and n-type doping such as the addition of Ni into the half Heusler structure.[1-5] In this work, n-type heterovalent substitution of half-metallic MnNiSb is made to semiconducting TiNiSn in the half Heusler structure. The constituents of TiNiSn and MnNiSb are low cost and earth-abundant materials which are of interest to industrial applications. In a half Heusler structure, admixing of a half-metal in a semiconductor is regarded promising due to the metallic inclusion in the matrix. This combination results in high phonon scattering and phase separation which lowers the thermal conductivity.[6] The 'phonon-glass / electron-crystal' concept is applicable in the $Ti_{1-x}Mn_xNiSn_{1-x}Sb_x$ system.[7] In this approach, there is high electron mobility due to the semiconducting-TiNiSn and n-type doping, phonon glass disorder by the half-metal, and plausibly phase separation without interfering with the electron path.

Mn and Sb are substituted in the Ti and Sn sites, respectably.

This substitution introduces electrons as charge carrier in the $Ti_{1-x}Mn_xNiSn_{1-x}Sb_x$ system. Mn is known to induce a phase separation in TiCoSb half Heusler materials, resulting in a lowering of the thermal conductivity and an improvement of the figure of merit.[8] The lowering of thermal conductivity by Mn is promising because half Heusler TEs usually have a relatively low figure of merit resulting in a high thermal conductivity. Also, Sb causes an increase in power factor when doped in the TiNiSn system.[9,10] Therefore, Mn and Sb are proposed to be promising dopants in the TiNiSn system.

The aim here is to understand the phase distribution and changes of the physical properties on doping electrons in both Ti and Sn sites. The half Heusler compounds crystallize in the cubic space group *F-43m* (216, MgAgAs structure type). The best TE properties were observed for x=0.05 with a *zT*=0.45 and x=0.02 with a power factor of 3.65, whereas x=0.1 showed a drastic deterioration in the properties of the quinary system. The main phase in most samples is TiNiSn with some Mn and Sb incorporation into the matrix. All samples had impurity phases that contributed some scattering and point defects. The characterization was done using powdered X-ray diffraction patterns, energy-dispersive X-ray spectroscopy, scanning electron microscopy, differential scanning Calorimetry. Electrical and thermal transport properties were measured from Helium temperatures up to 150 K.

## Experimental

The sample of the series $Ti_{1-x}Mn_xNiSn_{1-x}Sb_x$ (x= 0, 0.01, 0.02, 0.03, 0.05, and, 0.1) were prepared by arc melting in a purified argon atmosphere using a mixture of stoichiometric amounts of the half-Heusler compounds TiNiSn and MnNiSb. The as-cast samples were annealed at 973/1073 K for 7 days based on the differential thermal analysis (DTA) results. The annealing was followed by quenching in ice water. The samples were cut into rectangular bars 1.5x1.5x10 mm³ and squares 1.5x10x10 mm³, which were used later for characterization and thermoelectric properties measurements.

[a] Max Planck Institute for Chemical Physics of Solids, Nöthnitzer Str. 40, 01187 Dresden, Germany.
[b] Johannes Gutenberg University Mainz, Staudingerweg 9, 55128 Mainz.
*Email: Gerhard.Fecher@cpfs.mpg.de

The rectangular bars were used for both the low and high temperature Seebeck coefficient and electrical conductivity measurements. The rectangular bar was also used for the low temperature thermal conductivity and the square was used for the high temperature thermal conductivity measurements.

Powder X-ray diffraction (PXRD) measurements were performed with Cu Kα radiation at room temperature, using an image-plate Huber G670 Guinier camera equipped with a Ge(111) monochromator. The thermal analysis was done using differential scanning calorimeter NETZSCH model DSC 404 C Pegasus, under Ar atmosphere. The microstructures of the samples were examined by scanning electron microscopy (SEM) using a Philips X scanning electron microscope. Quantitative electron probe microanalysis (EPMA) of the phases was carried out by using an energy dispersive X-ray spectroscopy (EDX) analyser (Phoenix V 5.29, EDAX) and a wavelength-dispersive spectrometer WDXS (Cameca SX 100) with the pure elements as standards (the acceleration voltage was 25 kV, using the K- and L-lines). The grain sizes and phase percentages were determined from a back scattering electron (BSE) images using scanning electron microscope JSM 7800F. The software STREAM from Olympus-SIS was used to define the grain boundaries and the brightness intensity. The chemical analysis was made using the inductively coupled plasma-optical emission spectroscopy (ICP-OES) analysis using Agilent ICP-OES 5100 SVDV.

The low temperature (0 to 300 K) Seebeck coefficient, electrical conductivity, and thermal conductivity were measured using a physical properties measurement system (PPMS, Quantum Design). The high temperature (300-875 K) Seebeck coefficient and resistivity were measured using ULVAC ZEM-3 system. The high temperature (300-875 K) thermal diffusivity was determined using the laser flash analysis (LFA 457, NETZSCH). The thermal conductivity was determined using the equation $\kappa = \alpha \rho C_P$, where $\alpha$ is the thermal diffusivity, $C_P$ is the specific heat (measured using DSC, STA 449, NETZSCH), and $\rho$ is the density (measured using the (AccuPyc 1330 series No. 2441, Micromeritics Instruments under He-atmosphere). The carrier concentration was determined using Hall effective measurements (IPM-HT-Hall-900K system).

## Results and discussion

### Phase Characterization:

The phase characterization was done through PXRD, DSC, BSE, EPMA, and SEM and BSE images. The thermal analysis was used to estimate the annealing conditions. In Figure 1, the DSC curve of as cast $Ti_{0.98}Mn_{0.02}NiSn_{0.98}Sb_{0.02}$ is shown. The melting point of this specimen is 1720 K. All of the specimens had a similar melting point and were multiphase. Since each specimen had a slightly different combination of phases, DTA-DSC had to be studied to determine the specific annealing condition. In the case of $Ti_{0.98}Mn_{0.02}NiSn_{0.98}Sb_{0.02}$, the annealing conditions were chosen to be 1073K for 170 hours. Specific care was taken that the impurity melting points from the DSC measurement did not overlap or were close to the annealing temperature.

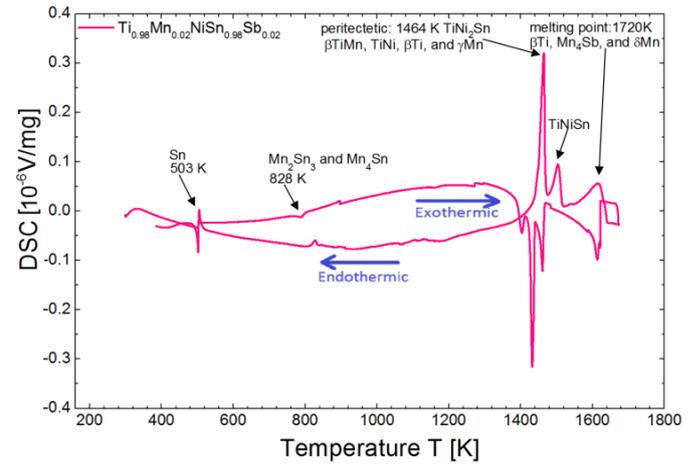

Fig. 1: Differential scanning calorimetry analysis for as cast $Ti_{0.98}Mn_{0.02}NiSn_{0.98}Sb_{0.02}$ system prepared by arc melting.

The diffraction patterns revealed that the samples were multiphase throughout the series, as seen in Figure 2. The majority phase was TiNiSn after annealing. The minority phase consists of $TiNi_2Sn$, Sn, and $Ti_6Sn_5$ with low reflection intensities. These minority phases are stable and are repeatedly seen in the intrinsic phase separation of the TiNiSn-type system through various sample preparation methods.[2,9,19] The lattice parameter does not show a pattern or significant amount of change through the series of doping, as seen in Table-1. The weak effect of the substitutions on the lattice parameter is because Ti is replaced with Mn and Sn is replaced with Sb. Ti has a larger atomic radius than Mn, but Sn has a smaller atomic radius than Sb.

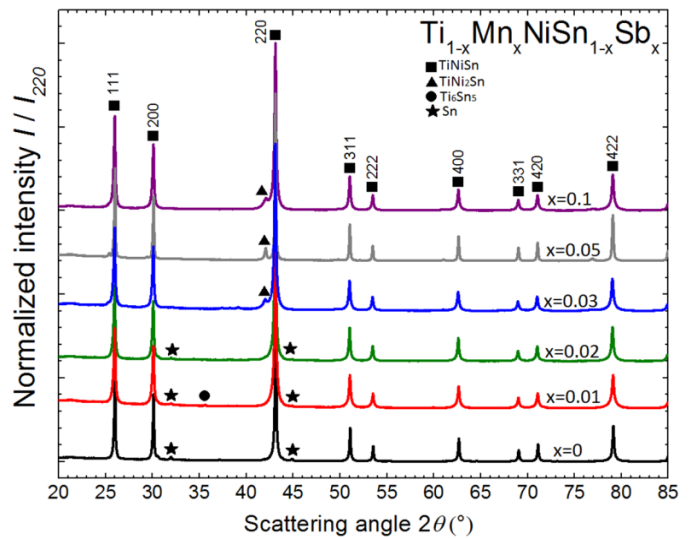

Fig. 2: X-ray diffraction patterns of $Ti_{1-x}Mn_xNiSn_{1-x}Sb_x$ (x= 0, 0.01, 0.02, 0.03, 0.05, and 0.1) alloys prepared by arc melting and then annealing at 1073/1173 K for 7 days.

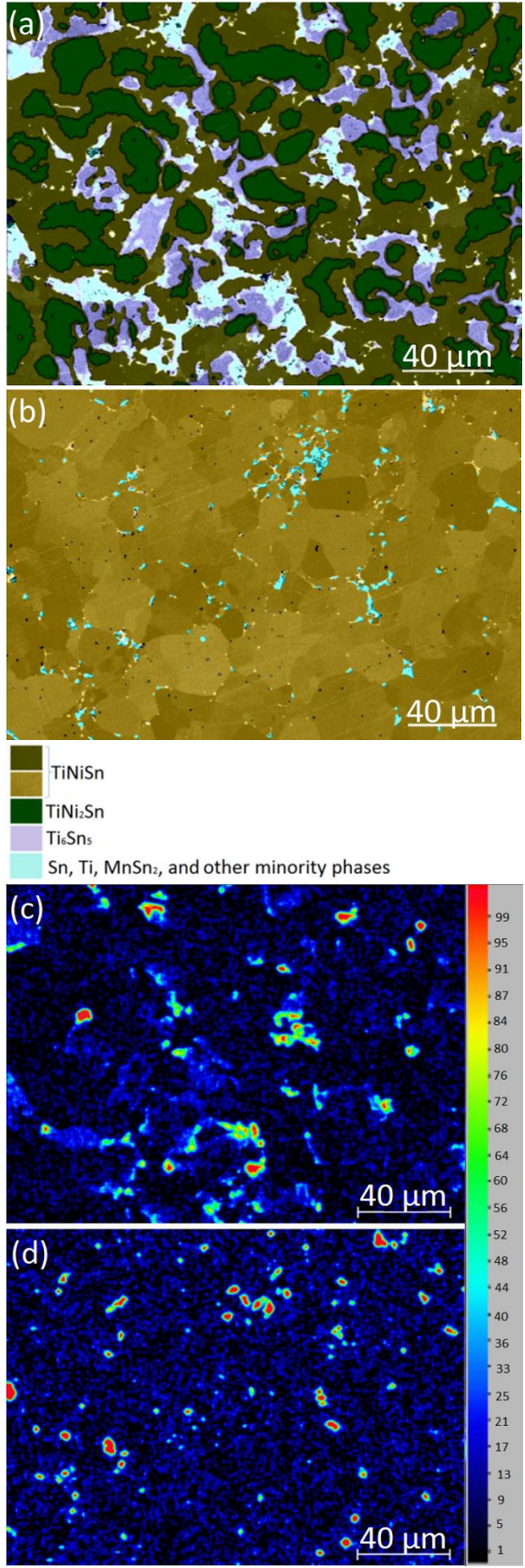

Fig. 3: Back scattering electrons SEM images in the $Ti_{0.98}Mn_{0.02}NiSn_{0.98}Sb_{0.02}$ system: (a) as cast specimen synthesized by arc melting, (b) post annealing specimen treated at 1073 K for 7 days, (c) Mn content in the as cast specimen, and (d) Mn content in the post annealing specimen.

This substitution causes the interatomic distances to show an inconsistent distribution, which then results in an irregular pattern in the lattice parameter. Several of the phases found from the PXRD patterns were in agreement with the phases found in the metallographic, which includes TiNiSn, $TiNi_2Sn$, and Sn.

Figure 3 shows the microstructure of $Ti_{0.98}Mn_{0.02}NiSn_{0.98}Sb_{0.02}$ alloys as-cast and after annealing. The as cast sample shows 19.4% deviation in the TiNiSn phase, however, this deviation is reduced to 1.6% after annealing. The large deviation in the as-cast material was due to inhomogeneity. The main phase TiNiSn had some Ni-rich areas which contributed to a broad spectrum of deviation. This Ni-rich main phase was not present in the annealed specimen, rather there were minority areas of Mn substitution in the Ti position, forming MnNiSn. Sb was seen in trace amounts in the main phase of both the as-cast and annealed specimen. Sb has a relatively low enthalpy of vaporization in comparison to other elements used in the alloy. Due to this fact, it was made sure that the percent yield after arc melting was 99% and in some cases, the chemical analysis was also made to confirm the overall composition. The phase separation is less in the annealed specimen in comparison to the as cast. The major reason for this change is because the grain size became larger upon annealing and some of the minority phases agglomerated. Also, Mn uniformly integrated into to the main phase after annealing in comparison to the as-cast, as seen in Fig-3(c,d).

| x | Lattice parameter [Å] | Grain size [μm] | Phase percentage (%) | | | Density [g/cm$^3$] |
|---|---|---|---|---|---|---|
| | | | Majority | Minority | Voids | |
| 0 | 5.9285 | 10.6 | 88.67 | 7.92 | 1.99 | 7.76 |
| 0.01 | 5.9370 | 10.4 | 96.95 | 1.35 | 0.00 | 7.46 |
| 0.02 | 5.9350 | 10.6 | 96.93 | 2.37 | 0.10 | 7.47 |
| 0.03 | 5.9360 | 11.3 | 95.73 | 0.45 | 0.01 | 7.70 |
| 0.05 | 5.9330 | 10.6 | 80.89 | 14.01 | 0.00 | 7.43 |
| 0.1 | 5.9312 | 10.2 | 85.67 | 7.79 | 5.30 | 7.78 |

Table 1: Lattice parameter, average grain size, phase percentage, and density in the $Ti_{1-x}Mn_xNiSn_{1-x}Sb_x$ system. Note, that the phase percentage was calculated using 100 x 100μm BSE-SEM images.

## Thermoelectric Properties
### Electrical Transport:

The electrical conductivity and Seebeck coefficient as a function of temperature are presented in Figure-3(a,b) and the 600 K values are stated in Table-2. The high and low-temperature data sets were measured using separate instruments. The two temperature sets do not have that much variation, so it confirms the stability of the material to a certain extent. However, the heating and cooling curves show some variation suggesting instability of the samples which might be due to the presence of certain minority phases.

| x | α [μV/K] | σ [$10^3$ S/m] | $α^2σ$ [$10^3$ W/Km] | κ [W/Km] | $κ_l$ [W/Km] | $κ_e$ [W/K*m] | zT | μ [$cm^2$/Vs] | n [$10^{21}$ $cm^{-3}$] | ms* [me] |
|---|---|---|---|---|---|---|---|---|---|---|
| 0 | -132.4 | 68 | 1.21 | 6.6 | 5.8 | 0.8 | 0.11 | 1777 | 0.24 | 1.36 |
| 0.01 | -176.8 | 91 | 2.66 | 6.2 | 5.2 | 1.0 | 0.25 | 2849 | 0.15 | 1.32 |
| 0.02 | -190.2 | 73 | 2.67 | 6.4 | 5.9 | 0.5 | 0.25 | 1294 | 0.25 | 2.02 |
| 0.03 | -151.5 | 120 | 2.77 | 7.6 | 6.3 | 1.3 | 0.22 | 1348 | 0.56 | 2.71 |
| 0.05 | -133.4 | 163 | 2.90 | 6.3 | 4.5 | 1.8 | 0.18 | 1153 | 0.56 | 2.39 |
| 0.1 | -54.6 | 403 | 1.20 | 13.7 | 9.9 | 3.8 | 0.05 | 429 | 5.87 | 4.70 |

Table 2: Seebeck coefficient (α), electrical conductivity (σ), power factor ($α^2σ$), thermal conductivity (κ), lattice thermal conductivity ($κ_l$), electronic thermal conductivity ($κ_e$), figure of merit (zT), hall mobility (μ), carrier concentration at (n), and Seebeck effective mass (ms*) for the $Ti_{1-x}Mn_xNiSn_{1-x}Sb_x$ system. Note that all the physical properties listed above were measured or calculated at 600 K with exception of the carrier concentration which was calculated at 300 K.

According to the DSC results, the specimens are multiphased. Certain minority phases such as Sn and $Mn_2Sn_3$ have lower melting points than the temperature range covered by the measurement. The variation in the heating and cooling could be because small areas in the sample are in the peritectic zone when it reaches higher temperatures.

The negative sign on the Seebeck coefficient confirms that the charge carriers are electrons. The Seebeck coefficient is the largest for x=0.02. The coefficient increases till x=0.05, but after that, it shows a downward trend. The alloys act as a semiconductor or a degenerate metal until x=0.05, but after that, it shows metallic character. From Table 2, we see that as the carrier concentration increases the Seebeck coefficient decreases throughout the series. The Seebeck coefficient is inversely proportional to carrier concentration for semiconductors and degenerate metals.[11] Other factors such as the Fermi energy level, density of states, and temperature gradient could also affect the Seebeck coefficient.[11,12] From Figure-3(a), it can be taken the overall nature of all except x=0.1 specimens is semiconducting because the Seebeck coefficient first increases and then decreases in a parabolic manner. The parabola gets less steep as the doping increases, which confirms a decline in the semiconducting properties. At x=0.1, the Seebeck coefficient varies rather linearly with temperature, corresponding the acquired metallic character.

Table 2 also reports the calculated Seebeck effective mass at 600 K. The Seebeck mass accounts for the density of states. This calculation was done assuming the scattering parameter is zero since this trend is common for acoustic phonon scattering above 300 K.[13] Table 2 shows an inverse relation for Seebeck mass and Hall mobility, this trend is usually seen in materials with a small band gap.[7] From the electronic properties trend, it is established that the nature of the material is semiconducting for x≤0.05. Therefore, the relationship between the Seebeck mass and Hall mobility is intact.

Electrical conductivity gradually increases as the amount of the half-metal, NiMnSb is increased. In accordance with the Seebeck coefficient trend, the electrical conductivity also supports the acquired metallic nature throughout the series. In Figure-3(c), x=0, 0.01, and 0.02 shows a metal like character for temperatures below 200 K, whereas at x>0.03 this metallic character subsides. It is well known that TiNiSn is a semiconductor.[1] Since the annealed TiNiSn is not single phase, there are impurities in the form of point defects and vacancies at the interstitial positions in the material. Frenkel defects cause some of these impurities are in the form of the full Heusler $TiNi_2Sn$ which is metallic in nature or $TiNi_{1+y}Sn$, where y≥0.1. Such Ni-rich phases were also seen in the metallographic characterization. In addition, the low temperature changes in x=0, 0.01, and 0.02 could also be due to a scattering mechanism which could affect the band structure leading to a crossover from metallic to semiconducting behavior.[14]

The carrier concentration is calculated from the Hall measurements at low temperatures. Figure 3(d) and Table 2 show the carrier concentration of all the investigated specimens, which mostly increases as the n-type substitution is increased. The carrier concentration for x=0 is higher than x=0.01 because TiNiSn was not entirely single phase there are some Ni-rich spots. In the absence of Mn, the TiNiSn system consists of some Ni-rich point defects and the scattering mechanism caused by the Frenkel defects that increases the overall carrier concentration.

**Thermal Transport:**

There is a considerable amount of discontinuity in the thermal conductivity between the low and high temperature measurements. The major cause of this discontinuity is that different instrumentations and sample specimens were used for the two temperature ranges. Different specimens of the same composition, may result in minor error, considering the difference in texture and grain size of the specimens. Another minor reason for the discontinuity could be because of the bipolar interactions of electrons and holes and the gradual increase in the electronic thermal conductivity.[15] At low temperatures, the thermal conductivity follows an expected trend in agreement with the electronic properties. Thermal conductivity shown in Figure-4(a) does not follow a simple trend at high temperature. At x=0.01 and 0.02 the thermal conductivity decreases, but at x=0.03 the thermal conductivity increases back again. There are several factors that may account to the disorder in the trend, such as mean free path and other intrinsic properties.

The total thermal conductivity is the sum of phonon ($\kappa_l$), electronic ($\kappa_e$), magnon ($\kappa_m$), and bipolar effect ($\kappa_{bi}$) contributions.[1,16,17] The Wiedemann-Franz law, is stated in equation (2) below

$$\kappa = \kappa_l + \kappa_e + \kappa_m + \kappa_{bi} \qquad (1)$$
$$\kappa_e = L\sigma T \qquad (2)$$

where $L$ is the Lorentz number, $\sigma$ is the electrical conductivity, and $T$ is the temperature.

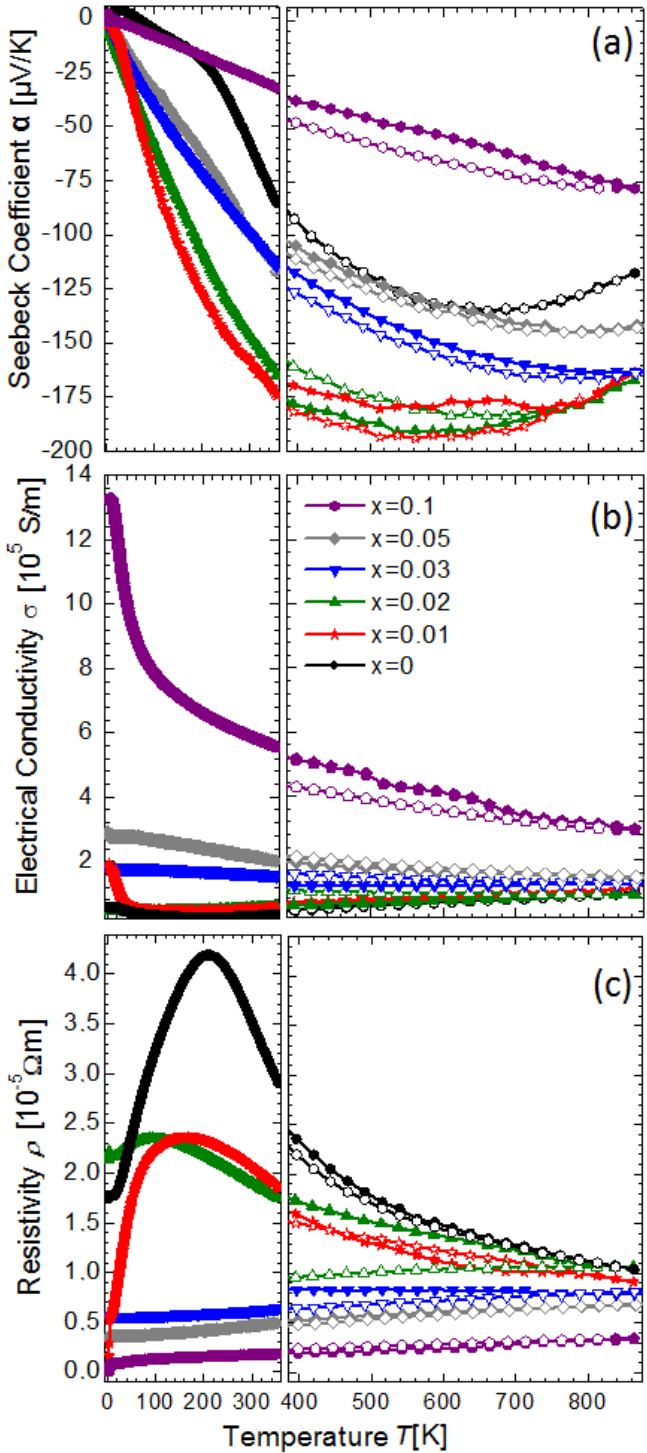

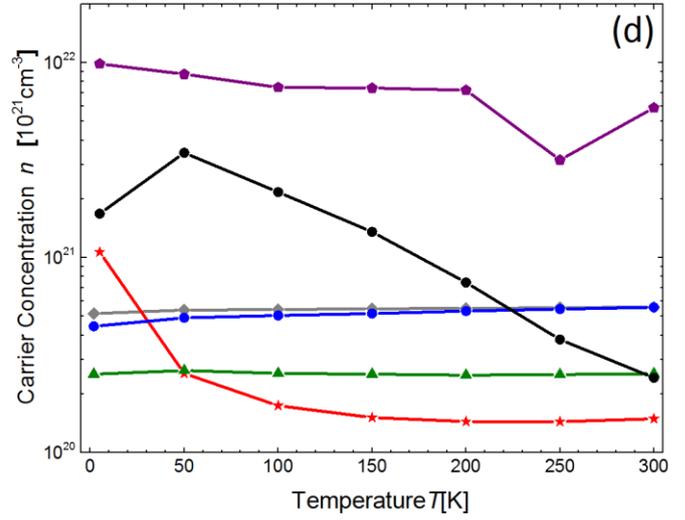

Fig. 4: Electronic properties as a function of temperature in the $Ti_{1-x}Mn_xNiSn_{1-x}Sb_x$ (x= 0, 0.01, 0.02, 0.03, 0.05, and 0.1) system: (a) Seebeck coefficient (b) electrical conductivity (c) resistivity, and (d) carrier concentration. Note that the filled symbols represent heating and the empty symbols represent cooling in (a), (b), and (c).

The magnon and bipolar effect are negligible in comparison to the lattice and electron term since the here investigated Heusler phases are non-magnetic.[1,16] Therefore, the lattice and electron term can be calculated individually using the above law. The Lorentz number was calculated using equation (3) below. This equation gives a better approximation than $L$=2.4x10$^{-8}$ WΩK$^{-2}$ because the Seebeck coefficient is taken into account.[18]

$$L = 1.5 + e^{\frac{|-\alpha|}{116}}, \text{ where } L \text{ is } 10^{-8} \text{ WΩK}^{-2} \qquad (3)$$

Table 2 and Figure 5 show that the lattice thermal conductivity is much larger than the electronic thermal conductivity. The electronic thermal conductivity increases through the series as n-type charge carriers are introduced into the system. The increase in the electronic thermal conductivity also confirms the acquired metallic character through the series. Figure-4(a) shows an increase in the low-temperature thermal conductivity from 0 to 100 K due to a decrease of acoustic phonon scattering and electron scattering. The lattice thermal conductivity is in close relation with the total thermal conductivity, however it does not show a simple pattern through the series. Lattice thermal conductivity is affected by various factors such as the point defects, phase separation, increasing molar mass by isoelectronic substitution, acoustic phonon effect, mean free path, and other extrinsic properties.[4,9,16,18-20] One of the reasons for the rather large lattice thermal conductivity could be because of large grain size after annealing as seen in Table-1. Materials with high phonon scattering typically have a high number of density of boundaries by reducing the grain size.[21]

The phonon mean free path is directly proportional to the phonon thermal conductivity. The mean free path can be calculated from equation (4) and the 600 K values are stated in Table-2.[22]

$$L_{ph} = \frac{3\kappa_{lattice}}{v_s C} \qquad (4)$$

Where $L_{ph}$ is the phonon mean free path, $v_s$ is the velocity of sound, and $C$ is the heat capacity. At 600 K, the mean free path increases in an irregular manner down the series. The mean free path can also be affected by Frenkel defect. The vacancies and point defects are specific to the intrinsic properties and phase separation.

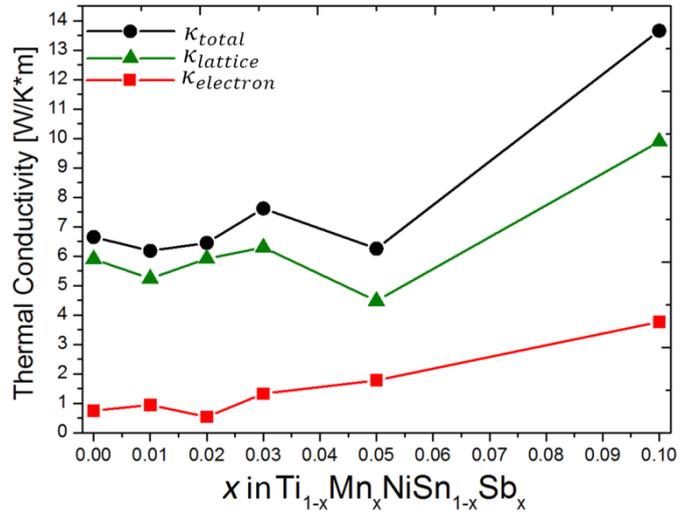

Fig. 5: Thermal conductivity: total, lattice, and electronic as a function of $x$ in the $Ti_{1-x}Mn_xNiSn_{1-x}Sb_x$ ($x$= 0, 0.01, 0.02, 0.03, 0.05, and 0.1) system at 600 K.

## Thermoelectric Performance

In figure 4(b), all the substituted specimens except x=0.1 showed an improvement in the *zT*. Here, the *zT* of TiNiSn is a little lower than that reported in the literature. Since the samples were prepared entirely through arc-melting and annealing, the overall figure of merit is not very high. Other preparation methods such as ball milling, microwave, and spark plasma sintering have reported much higher *zT* for TiNiSn.[4,20,23] Therefore, the potential of having higher performance through other preparation techniques is high. The discrepancy in the high and low temperature measurements of $\kappa$ causes a minor misalignment in the *zT* graph.

The power factor is maximum for x=0.02 and 0.03 at 600 K, as seen in Figure-4(c). Interestingly, the power factor of x=0.1 was greater than x=0 specimen because of its amplified electrical conductivity on acquiring a metallic character. These materials also surpass the power factor of materials with Zr and Hf substitution in Ti-site.[24] The high lattice thermal conductivity counteracts the power factor, which lowers the overall figure of merit. Moreover, he investigated specimens had considerably low mechanical stability on heat treatment, which would explain the discrepancies in the high and low temperature measurements and the raise in thermal conductivity.

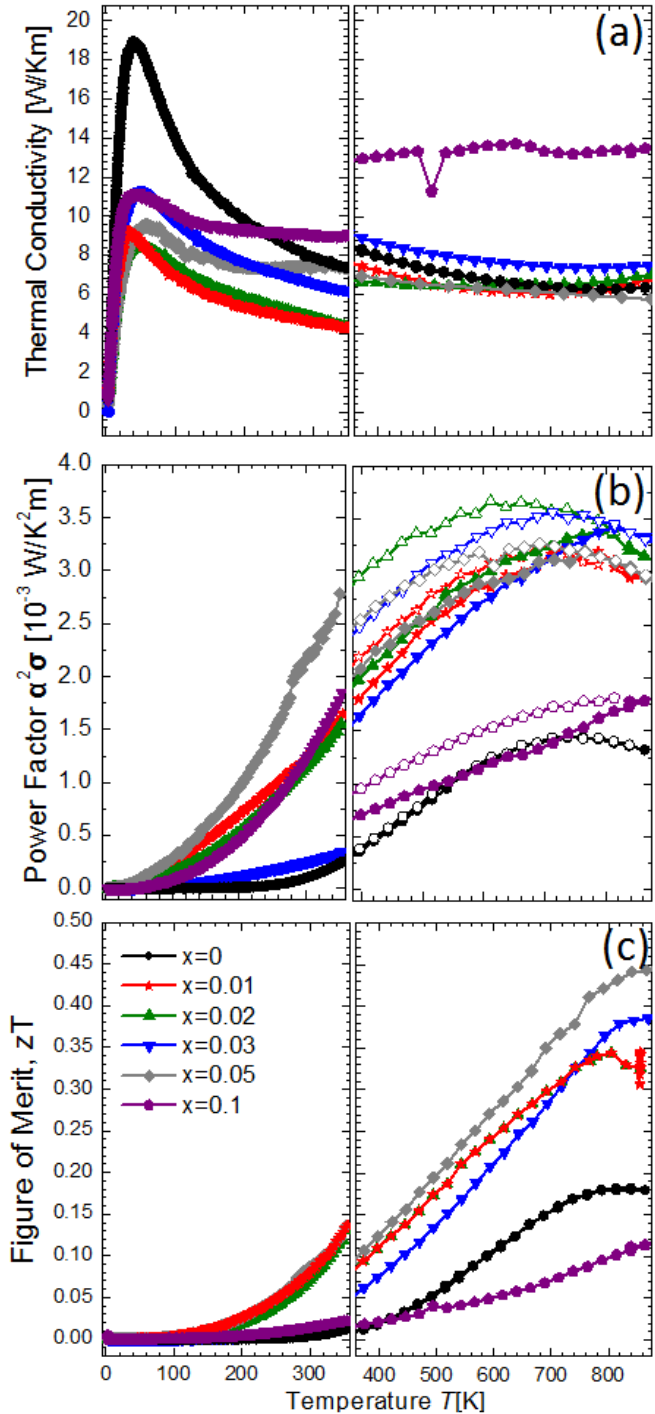

Fig. 4: Thermal transport properties and thermoelectric performance as a function of temperature in the $Ti_{1-x}Mn_xNiSn_{1-x}Sb_x$ (x= 0, 0.01, 0.02, 0.03, 0.05, and 0.1) system: (a) thermal conductivity, (b) power factor, and (c) figure of merit, zT. Note that the filled symbols represent heating and the empty symbols represent cooling in (b) power factor.

## Conclusion

Samples of the $Ti_{1-x}Mn_xNiSn_{1-x}Sb_x$ system were synthesized by arc melting and annealing at 1073/ 1173 K, and their phase distribution and TE properties were studied. The microstructure after annealing showed some Ni-rich TiNiSn and Mn-Sn binary impurities. Ni-rich TiNiSn were shown to be caused by Frenkel defects. The transport properties confirmed a crossover by from metallic to semiconducting behavior from 0-100 K temperature range as a result of the defect. The overall figure of merit showed a 1.6 fold increase from x=0 to x=0.05. The alloys showed semiconducting behavior from 0≤x≤0.05 and becomes a degenerate metal at x=0.1. The power factor is the highest for x=0.02 (3.65x10$^{-3}$ W/Km at 650 K), which is larger than corresponding alloys with expensive dopants such as Hf and Zr in the Ti-site.[24] The Seebeck coefficient is also considerably enhanced (α=-191 µV/K at 540 K) for x=0.01 and 0.02. In total, the thermoelectric performance of TiNiSn is drastically enhanced by mixing TiNiSn with MnNiSb in the half Heusler structure. Therefore, the Ti$_{1-x}$Mn$_x$NiSn$_{1-x}$Sb$_x$ system can be characterized as a promising thermoelectric in the sunstitution range x=0.01-0.05.

## Acknowledgements

We thank S. Kostmann for specimen preparation for the microstructural examinations, M. Eckert and P. Scheppan and Dr. U. Burkhardt for SEM/EDX investigations, and R. Koban for transport measurements.

## References

1 S. Ouardi, G. H. Fecher, B. Balke, X. Kozina, G. Stryganyuk, C. Felser, S. Lowitzer, D. Ködderitzsch, H. Ebert and E. Ikenaga, Electronic Transport Properties of Electron and Hole-Doped Semiconducting C1b Heusler Compounds: NiTi1xMxSn (M = Sc, V), *Phys. Rev. B: Condens. Matter Mater. Phys.*, 2010, 82, 085108. 46
2 D. Jung, K. Kurosaki, C. Kim, H. Muta and S. Yamanaka, Thermal Expansion and Melting Temperature of the Half-Heusler Compounds: MNiSn (M = Ti, Zr, Hf), *J. Alloys Compd.*, 2010, 489, 328–331.
3 S. Sakurada and N. Shutoh, Effect of Ti Substitution on the Thermoelectric Properties of (Zr,Hf)NiSn Half-Heusler Compounds, *Appl. Phys. Lett.*, 2005, 86, 082105.
4 C. S. Birkel, J. E. Douglas, B. R. Lettiere, G. Seward, N. Verma, Y. Zhang, T. M. Pollock, R. Seshadri, and G. D. Stucky, Improving the thermoelectric properties of half-Heusler TiNiSn through inclusion of a second full-Heusler phase: microwave preparation and spark plasma sintering of TiNi$_{1+x}$Sn, *Physical Chemistry Chemical Physics*, 2013, *15*, 6990-6997.
5 R. A. Downie, D. A. MacLaren, R. I. Smith and J. W. G. Bos, Enhanced Thermoelectric Performance in TiNiSn-Based Half-Heuslers, *Chem. Commun.*, 2013, 49, 4184–4186.
6 M. Koehne, T. Graf, H.J. Elmers and C. Felser, *US Pat.*, 0156636, 2013.
7 G. J. Synder and E. S. Toberer, *Nat. Mater.*, 2008, 7, 105
8 T. Graf, P. Klaer, J. Barth, B. Balke, H.-J. Elmers and C. Felser, *Scr. Mater.*, 2010, 63, 1216–1219
9 S. Bhattacharya, A. L. Pope, R. T. L. Iv, T. M. Tritt, V. Ponnambalam, Y. Xia and S. J. Poon, Effect of Sb Doping on the Thermoelectric Properties of Ti-Based Half-Heusler Compounds, TiNiSn$_{1-x}$Sb$_x$, *Appl. Phys. Lett.*, 2000, 77, 2476–2478.
10 S. Chen and Z. Ren, *Mater. Today*, 2013, 16, 387
11 S. M. Kauzlarich, S. R. Brown and G. J. Snyder, *Dalton Trans.*, 2007, 21, 2099.
12 A. Kawamura, H. Yufei, S. M. Kauzlarich, Synthesis and Thermoelectric Properties of the YbTe-YbSb System, *Journal of Electronic Materials*, 2016, 45, 11664-015-4202.
13 Y. Tang, Z. M. Gibbs, L. A. Agapito, G. Li, H. Kim, M. Nardelli, S. Curtarolo and G. J. Snyder, *Nat. Mater.*, 2015, 14, 1223–1228.
14 C. Uher, J. Yang, S. Hu, D. T. Morelli and G. P. Meisner, *Phys. Rev. B: Condens. Matter Mater. Phys.*, 1999, 59, 8615.
15 M. Wagner, R. Cardoso-Gil, N. Oeschler, H. Rosner and Yu. Grin, *J. Mater. Res.*, 2011, 26, 1886.
16 B. L. Huang and M. Kaviany, *J. Appl. Phys.*, 2006, 100, 123507.
17 H. Alam and S. Ramakrishna, *Nano Energy*, 2013, 2, 190–212.
18 H.-S. Kim, Z. M. Gibbs, Y. Tang, H. Wang and G. J. Snyder, *APL Mater.*, 2015, 3, 041506
19 R. A. Downie, D. A. MacLaren, R. I. Smith and J. W. G. Bos, Enhanced Thermoelectric Performance in TiNiSn-Based Half-Heuslers, *Chem. Commun.*, 2013, 49, 4184–4186.
20 J. E. Douglas, C. S. Birkel, N. Verma, V. M. Miller, M.-S. Miao, G. D. Stucky, T. M. Pollock and R. Seshadri, Phase Stability and Property Evolution of Biphasic Ti–Ni–Sn Alloys for Use in Thermoelectric Applications, *J. Appl. Phys.*, 2014, 115, 043720.
21 S. Chen and Z. Ren, *Mater. Today*, 2013, 16, 387.
22 T. M. Tritt and M. A. Subramanian, *MRS Bull.*, 2006, 31, 188.
23 C. S. Birkel, W. G. Zeier, J. E. Douglas, B. R. Lettiere, C. E. Mills, G. Seward, A. Birkel, M. L. Snedaker, Y. Zhang, G. J. Snyder, T. M. Pollock, R. Seshadri and G. D. Stucky, *Chem. Mater.*, 2012, 24, 2558–2565
24 E. Rausch, B. Balke, S. Ouardi and C. Felser, *Phys. Chem. Chem. Phys.*, 2014, 16, 25258–25262